\documentclass[%
 reprint,
10pt,
showpacs,preprintnumbers,
 amsmath,amssymb,
 aps,
 pra, 
]{revtex4-1}

\usepackage{graphicx}
\usepackage{dcolumn}
\usepackage{bm}
\usepackage{hyperref}
\usepackage{color}
\usepackage[normalem]{ulem}

\newcommand{\ket}[1]{|{#1}\rangle}

\newcommand{\bb}[1]{\left(#1\right)}

\newcommand{\No}{\hat{N}_0}

\graphicspath{{./}}

\begin{document}

\title{Metrologically useful states of spin-1 Bose condensates with macroscopic magnetization}

\author{Dariusz Kajtoch,$^{1}$ Krzysztof Paw\l{}owski$\,^{2}$ and Emilia Witkowska$^{1}$}
\affiliation{$^{1}$Institute of Physics, PAS, Aleja Lotnikow 32/46, PL-02668 Warsaw, Poland\\
$^{2}$Center for Theoretical Physics, PAS, Aleja Lotnikow 32/46, PL-02668 Warsaw, Poland}

\date{\today}

\begin{abstract}
We study theoretically the usefulness of spin-1 Bose condensates with macroscopic magnetization in a homogeneous magnetic field for quantum metrology. 
We demonstrate Heisenberg scaling of the quantum Fisher information for states in thermal equilibrium. 
The scaling applies to both antiferromagnetic and ferromagnetic interactions. 
The effect preserves as long as fluctuations of magnetization are sufficiently small.
Scaling of the quantum Fisher information with the total particle number is derived within the mean-field approach in the zero temperature limit and exactly in the high magnetic field limit for any temperature. 
The precision gain is intuitively explained owing to subtle features of the quasi-distribution function in phase space.
\end{abstract}


\maketitle

\section{Introduction}

Atom interferometry techniques are widely used in most precise measurements of physical parameters e.g. time, force or strength of field.
The physical quantities are typically mapped onto a phase $\theta$ which is estimated afterwards.
The best precision in the $\theta$ estimation is limited by the Cram\'er-Rao bound $ \Delta \theta \gtrsim {F_{Q}}^{-1/2}$~\cite{braunstein1994}, where $F_{Q}$ is the quantum Fisher information (QFI).
The QFI quantifies potential improvement in the precision of the $\theta$ estimation. 
The scaling of the QFI with the total atom number $N$ is crucial.
When uncorrelated atoms are used, the best precision is given by the standard quantum limit (SQL) with $ F_Q\sim N$ originating from the statistical nature of quantum noise.
The SQL can be overcome using quantum resources such as squeezing and entanglement, potentially approaching the ultimate Heisenberg limit $ F_Q\sim N^2$. 
The QFI, by scaling with the atom number, fully identifies a class of useful for interferometry entangled states, and in this sense it is a witness for genuinely multi-particle entanglement \cite{PhysRevA.85.022321,PhysRevA.85.022322,Smerzi}.
The QFI is also geometrically interpreted as statistical speed of changes of a state subjected to an interferometer under an infinitesimal increment of $\theta$ \cite{PhysRevLett.72.3439}.
The QFI is optimized over all possible measurements used to estimate the phase. 
In the case of a concrete measurement, the notion of the classical Fisher information (FI) is used \cite{fisher_1925}.
The geometrical interpretation of the QFI provides an experimental tool to extract the FI \cite{Strobel2014}. 
Additionally, the FI is related to dynamical susceptibilities \cite{NaturePhys12} which can also be measured in experiments.

Multi-component Bose-Einstein condensates (BECs) of ultracold atoms have been already recognized as realistic, highly controllable and tunable systems for the entangled states generation 
\cite{{Strobel2014},{Gross2010},{Riedel2010},{PhysRevLett.111.180401},{PhysRevA.93.033608},{PhysRevA.93.023627},{RevModPhys.85.1191},{NaturePhys8}} useful for atomic interferometry \cite{vinit2017, Ockeloen2013}.
In this work, we concentrate on antiferromagnetic and ferromagnetic spin-1 Bose condensates that constitute three-mode systems numerated by the quantum magnetic number $m_F=0,\pm 1$ for the situation of current experimental relevance \cite{RevModPhys.85.1191}, where $N$ is large and the magnetization of the system $M$ (population difference between the extreme Zeeman levels) is conserved. 
Usefulness of the system for atomic interferometry is investigated experimentally nowadays \cite{{NaturePhys8},{PhysRevLett.115.163002},{10.2307/41351684},{TFScience2017}}, including observation of twin-Fock states~\cite{{10.2307/41351684},{TFScience2017}}, however for zero magnetization. 

The spin-1 Bose condensates have been also extensively studied due to the presence of the quantum phase transition at the critical value of an external magnetic field \cite{LiuLett2009, David2012, ChapmanPNAS, PhysRevLett.116.155301, PhysRevA.94.043635, PhysRevLett.110.045303, PhysRevLett.99.130402, PhysRevA.95.053638}.
The order parameter can be e.g. the occupation of the $m_F=0$ mode.
The quantum phase transition can be probed with the QFI \cite{NaturePhys12}.
The QFI of pure states is proportional to the variance of a rotation generator which may be estimated by the variance of an order parameter.
As a consequence, in the system we are interested in, one can expect from \cite{gerbier2013} that indeed at zero temperature the QFI is sensitive to the phase transition, and it has to be $O(N^2)$ for the quantum singlet state that arises in a vicinity of the critical point.

Our primary interest was to study the quantum phase transition in the system at non-zero temperature using the QFI. Unexpectedly, by performing numerical calculations of the QFI within the exact diagonalization method, we found that thermal states of the system with macroscopic magnetization, i.e. $M=O(N)$ and $N-M = O(N)$, have the Heisenberg scaling of the QFI not only in a surrounding area of the critical point and at zero temperature but in a broad range of magnetic fields, temperatures and interaction strengths. 
Thermal states of the system provide Heisenberg scaling of the QFI as long as the variance of magnetization is below $1$. 
The resulting scaling can be intuitively understood by considering the Wigner~\cite{Agarwal1994} quasi-distribution, as explained in Sec. \ref{sec:T0}.
We derived analytically scaling of the QFI with the total atom number $N$ using the mean-field approach for zero temperature and in the high magnetic field limit (HMFL) for any temperature, showing that the SQL can be still circumvented if the variance of magnetization is smaller than~$N$.  
Our results show that the states with macroscopic magnetization of reduced variance are immensely useful for interferometric purposes for a wide range of magnetic fields and interaction strengths.

\section{The system and the model} 
The system under consideration is a dilute spin-1 BEC in a homogeneous magnetic field with all three Zeeman states numerated by the magnetic quantum numbers $m_F=0,\pm 1$. 
We assume that the single mode approximation (SMA) is valid and all atoms share the same spatial wavefunction $\phi(\mathbf{r})$ \cite{PhysRevLett.81.5257, PhysRevA.66.011601} normalized to 1.  
The Hamiltonian that governs the spin dynamics is \cite{PhysRevA.82.031602}
\begin{equation}\label{eq:single_mode_ham}
\frac{\hat{\mathcal{H}}}{\tilde{c}} = \frac{{\rm sign}(c_2)}{2 N}\hat{J}^2 - q\hat{N}_{0},
\end{equation}
where $\hat{J}^2$ is the total spin operator and $\hat{N}_{m_F}$ is the particle number operator for the Zeeman state $m_F$. The energy unit is $\tilde{c} = N |c_2| \int d^3{r}|\phi(\mathbf{r})|^4$, where
$c_2=4\pi\hbar^2(a_2 - a_0)/3\mu$,  $\mu$ is an atomic mass, and $a_0$ and $a_2$ are the s-wave scattering lengths for two spin-$1$ atoms colliding in the combined symmetric channel, respectively, of spin $0$ and $2$ \cite{PhysRevA.66.011601}. For $c_2 < 0$ the interaction term alone favors the ferromagnetic phase (e.g. rubidium-87), whereas for $c_2 > 0$ the antiferromagentic phase (e.g. sodium-23). The second term in~\eqref{eq:single_mode_ham} describes the quadratic Zeeman energy, where $q=Q/\tilde{c}$ and $Q=(\mu_B {\cal B})^2/(4 E_{\rm hf})$
depends on the magnetic field strength ${\cal B}$, the Bohr magneton $\mu_B$ and the hyperfine energy splitting $E_{\rm hf}$ which can be both positive and negative \cite{PhysRevA.73.041602, TFScience2017} (see Appendix \ref{apx:paramters} for corresponding physical parameters). 
The total number of atoms operator $\hat{N} = \sum_{m_F} \hat{N}_{m_F}$ and the operator $\hat{J}_z =   \hat{N}_{+1}- \hat{N}_{-1}$ are both conserved. Terms proportional to $\hat{N}$ and $\hat{J}_z$ have no influence on the results, and they are dropped in the final form of (\ref{eq:single_mode_ham}).

We assume that the system is in an incoherent mixture of states which are thermal in the sector of fixed magnetization ~\cite{Corre2015}:
\begin{equation}\label{eq:rho_state}
\hat{\rho} = \sum\limits_{M=-N}^{N} w_{M}\hat{\rho}_{M},
\end{equation}
where $\hat{\rho}_{M} =e^{-\beta \hat{\mathcal{H}}}/\mathcal{Z}_{M}$ is a thermal state in the subspace of fixed magnetization $M = \langle \hat{J}_z\rangle$ and $\mathcal{Z}_{M}$ is the partition function in this subspace. The temperature $T$ is controlled by the parameter $\beta = \tilde{c}/(k_B T)$, where $k_B$ is the Boltzmann constant. The distribution of magnetization is given by the non-thermal weights $w_M = \exp[-(M -\bar{M})^2/2\sigma^2]/Z$, where $Z = \sum_{M} \exp[-(M -\bar{M})^2/2\sigma^2]$ and $\bar{M}$ is an average magnetization.
Notice, although the states $\hat{\rho}_{M}$  are thermal within the sector of fixed magnetization, the state $\hat{\rho}$ given by \eqref{eq:rho_state} is not thermal globally. 
As pointed out in ~\cite{Corre2015}, the state \eqref{eq:rho_state} describes states available in experiments \cite{ChangChapman2004, SadlerKurn2006, LiuLett2009, Guzman2011, David2012, PhysRevA.93.023614} better than the thermal globally state. 

The experiment we keep in mind is \cite{PhysRevA.93.023614}, where with appropriate techniques $\Delta^2 M$ may be made much smaller than in the case of the thermal fluctuations, but because of  small energy gap between internal levels, many of them are occupied within allowed sector of magnetizations.

\section{Interferometry \label{sec:Three-mode interferometry}}

\begin{figure}[t]
\includegraphics[width=\linewidth]{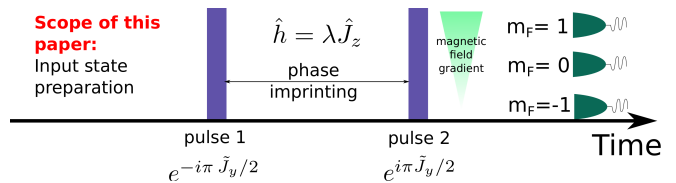}
\caption{The scheme of typical interferometric sequence. 
The input state is rotated with a pulse 1 to be sensitive to the Hamiltonian $\hat{h}\propto \lambda \hat{J}_z$. After the interrogation time, the value of the parameter to estimate $\lambda$ is encoded in the relative phases between paths of the interferometer. 
The pulse 2 is applied to map the phase onto quantities easy to measure.
The scheme describes as well the typical two arms Mach-Zehnder interferometer as the case with three arms, discussed in this paper.}
\label{fig:interf}
\end{figure}

The purpose of the paper is to identify quantum states, arising naturally in spin-1 Bose-Einstein condensates, which can be used to improve precision of measurements by using atom interferometry techniques. To make our results more clear, we first explain a possible application of spin-1 condensates in atomic interferometry.

In Fig. \ref{fig:interf} we illustrate a typical interferometric sequence. 
The initial state is splitted by a pulse ("pulse 1" on the diagram), evolving then separately and gaining the phase difference $\theta$. 
The source of the phase difference varies between interferometers. It may be caused either by the interaction with magnetic or gravitational fields, or by the internal energy difference between paths as in atomic clocks. 
We assume that accumulation of this phase difference is a dynamical process during which the evolution is governed by a linear Hamiltonian of the form $\hat{h} = \lambda \hat{J}_z$, where $\lambda$ is the parameter to estimate.
After the phase accumulation, the two paths are mixed by a second pulse ("pulse 2" on the diagram) and interfere.
The output state $\hat{\rho}_{\rm out}$ of the interferometer may be written as an input state $\hat{\rho}_{\rm inp}$ rotated by the angle $\theta$, precisely 
$\hat{\rho}_{\rm out} = U \hat{\rho}_{\rm inp} U^\dagger$, where $U=e^{-i \pi \tilde{J}_y/2}e^{-i \theta \hat{J}_z }e^{i \pi \tilde{J}_y/2}=e^{ -i \theta \tilde{J}_x}$ and $\theta = \lambda t$. Below, we use the word interferometer also to name the axis around which the state is rotated, so in this example the interferometer is $\tilde{J}_x$.
After the second pulse, one performs measurements, estimates the accumulated phase difference $\theta$ from the outcome, and then estimates the parameter $\lambda$.
From the Cram\'er-Rao bound, we know that the  uncertaintity of the parameter $\lambda$ estimation is limited from below by $1/\sqrt{F_Q[\hat{\rho}_{\rm inp}] }$. Consequently, the way to obtain high precision in the $\lambda$ estimation may be the use of an entangled state with a large value of the QFI. 
To benefit from the high value of QFI in metrology, one has to find such an observable which mean will be sensitive to rotations imposed by the phase accumulation. 
In general, it may be that there is no easy-to-measure observable making the state very difficult to use. This is the case for the cat states: to use them in the interferometer one has to measure a parity.

In this paper, we study theoretically the utility of the thermal states \eqref{eq:rho_state} of spin-1 Bose-Einstein condensates in the interferometer input.
As we will show later the state we are discussing are in close analogy to the Fock states. It is known, and it was already experimentally verified~\cite{{10.2307/41351684},{TFScience2017}}, that to benefit from them one has to measure second moment of the spin $\langle J_z^2 \rangle$. So by analogy, we expect that the states we discuss, would also require only measurements of the second moment of the collective spin.

The direct application of these states in atomic interferometry is not covered in more details by this paper.
Here, by using the QFI as a witness of the utility \cite{PhysRevA.85.022321}, we only identify good candidates among the states of the form \eqref{eq:rho_state}.
The system we are interested in can be used to measure e.g. the magnetic field, as its internal states differ with the quantum magnetic numbers.
In order to apply the states  \eqref{eq:rho_state} to the interferometer one should first erase the interaction and quadratic Zeeman terms in the Hamiltonian \eqref{eq:single_mode_ham}. It is usually done by making clouds of atoms more dilute, so that the number of two-body collisions drops down, and by switching off the external magnetic field used for the state preparation. 
The sample of the state \eqref{eq:rho_state} proposed in this paper should be rotated first \cite{Sanders1999-ws} and then placed in an unknown magnetic field that is sufficiently weak, so the quadratic Zeeman shifts would be of no importance~\cite{Ockeloen2013}. Then the value of the magnetic field would be, via the physical process corresponding to the Hamiltonian $\hat{h} \propto B \hat{J}_z$, imprinted in the phase difference between the $m_F=+1$ and $m_F=-1$ modes. The interferometer would follow the steps depicted in Fig. \ref{fig:interf}, with the appropriate choice of the operators $\tilde{J}_y$, $\tilde{J}_x$. In general, $\tilde{J}_y$ and $\tilde{J}_x$ are a linear superposition of $SU(3)$ algebra generators for our spin-1 system, and they are fully identified in the next section.

\section{Quantum Fisher Information}
The definition of the QFI follows from quantum estimation theory. The output state $\hat{\rho}_{\rm out}$ of any linear 3-mode interferometer, with equal phase difference $\theta$ between neighboring paths, can be written as $\hat{\rho}_{\rm out} =e^{-i \theta \hat{\Lambda}_{\mathbf{n}}} \hat{\rho} e^{i \theta \hat{\Lambda}_{\mathbf{n}}}$, where $\hat{\rho}$ is the input density matrix and $\hat{\Lambda}_{\mathbf{n}}=\mathbf{\hat{\Lambda}} \cdot \mathbf{n}$ where $\mathbf{\hat{\Lambda}} = \{ \hat{J}_x, \hat{Q}_{zx}, \hat{J}_y, \hat{Q}_{yz}, \hat{D}_{xy}, \hat{Q}_{xy}, \hat{Y}, \hat{J}_z \} $ is the vector of generators (see Appendix \ref{apx:generators} for definitions) spanning the bosonic $SU(3)$ Lie algebra and $\mathbf{n}$ is a unit vector defining the interferometer. The QFI is~\cite{Smerzi}
\begin{equation}\label{eq:fisher_information}
F_{Q}[\hat{\rho}, \hat{\Lambda}_{\mathbf{n}}] = 4\mathbf{n}^{T} \cdot \Gamma[\hat{\rho}] \cdot \mathbf{n},
\end{equation}
with the covariance matrix $\Gamma[\hat{\rho}]$ defined as
\begin{align}\label{eq:covariance}
\Gamma_{i,j}[\hat{\rho}] & = \sum\limits_{k} v_k \left[ \frac{1}{2}\langle k
|\{\hat{\Lambda}_i,\hat{\Lambda}_j \} | k \rangle - \langle k|
\hat{\Lambda}_i |k \rangle \langle k| \hat{\Lambda}_j |k\rangle \right] \nonumber\\
& -
4\sum\limits_{k>l}\frac{v_k v_l}{v_k + v_l}\text{Re}\left[\langle k|
\hat{\Lambda}_i |l\rangle \langle k| \hat{\Lambda}_j |l\rangle\right],
\end{align}
where we used eigenvectors and eigenvalues of the input state $\hat{\rho} = \sum v_k |k\rangle\langle k|$.
The maximal value of the QFI is given by the largest eigenvalue $\lambda_{\rm max}$ of the covariance matrix, $F_Q = 4\lambda_{\rm max}$. 
The optimal interferometer $\hat{\Lambda}_{\mathbf{n}}$ is defined by an eigenvector corresponding to $\lambda_{\rm max}$. 
Any separable state gives $F_Q \leqslant 4N$, while the maximal possible value $F_Q = 4N^2$ occurs only if the quantum state is fully particle entangled, \cite{PhysRevLett.96.010401,Smerzi}.

Although for eight generators the covariance matrix~\eqref{eq:covariance} has 36 different entries, its form simplifies significantly due to symmetries of $\hat{\rho}$ defined in Eq.\eqref{eq:rho_state}. The argument is the following.
Physical quantities such as the covariance matrix do not depend on the representation of the Hilbert space.
The density matrix $\hat{\rho}$ given in Eq. \eqref{eq:rho_state} commutes with $\hat{J}_z$. This implies that all  eigenstates $|k\rangle$ of the system's density matrix $\hat{\rho}$  have the following property:
\begin{equation}
e^{-i\varphi \hat{J}_z}|k\rangle = e^{-i\varphi M}|k\rangle,
\end{equation}
i.e. rotation of any eigenstate around the $\hat{J}_z$ operator results in a phase factor, which means that
$\Gamma[\hat{\rho}] = \Gamma\left[\hat{\rho}_{\varphi}\right]$, with
$\hat{\rho}_{\varphi} = e^{-i\varphi \hat{J}_z} \hat{\rho} e^{i\varphi \hat{J}_z}$.
On the other hand, one can rotate $SU(3)$ generators in definition (\ref{eq:covariance}) rather than the density matrix operator 
\begin{equation}
\Gamma[\hat{\rho}_{\varphi}] = M_{\varphi} \cdot \Gamma[\hat{\rho}] \cdot M_{\varphi}^T.
\end{equation}
For the order of generators defined by $\mathbf{\hat{\Lambda}}$ the rotation matrix $M_{\varphi}$ is given by
\begin{equation}
\left(\begin{array}{cccccccc} 
 \cos\varphi & 0 & -\sin\varphi & 0 & 0 & 0 & 0 & 0 \\
 0 & \cos\varphi & 0 & -\sin\varphi & 0 & 0 & 0 & 0 \\
 \sin\varphi & 0 & \cos\varphi & 0 & 0 & 0 & 0 & 0 \\
 0 & \sin\varphi & 0 & \cos\varphi & 0 & 0 & 0 & 0 \\
 0 & 0 & 0 & 0 & \cos2\varphi & -\sin2\varphi & 0 & 0 \\
 0 & 0 & 0 & 0 & \sin2\varphi & \cos2\varphi & 0 & 0 \\
 0 & 0 & 0 & 0 & 0 & 0 & 1 & 0 \\
 0 & 0 & 0 & 0 & 0 & 0 & 0 & 1
\end{array}\right).
\end{equation}
Combining everything we get the condition for the covariance matrix:
\begin{equation}
\forall_{\varphi}\ \ \ 
\Gamma[\hat{\rho}] = M_{\varphi} \cdot \Gamma[\hat{\rho}] \cdot M_{\varphi}^T.
\end{equation}
The above condition together with the definition of the covariance matrix, and because the Hamiltonian is self-adjoint, results in the following relations among non-zero elements left $\Gamma_{11} = \Gamma_{33}$, $\Gamma_{55} = \Gamma_{66}$, $\Gamma_{23} = -\Gamma_{14}$, $\Gamma_{22} = \Gamma_{44}$, $\Gamma_{34} = \Gamma_{12}$, and $\Gamma_{77} \neq 0$. 
Finally, the covariance matrix takes the block diagonal structure
\begin{equation}
\Gamma[\hat{\rho}] = \Gamma_0 \oplus \Gamma_0 \oplus [\Gamma_{55}] \oplus [\Gamma_{55}] \oplus [\Gamma_{77}] \oplus [0],
\end{equation}
where 
\begin{equation}
\Gamma_0 = \left( \begin{array}{cc} \Gamma_{11} & \Gamma_{12} \\ \Gamma_{12} & \Gamma_{22} \end{array}\right).
\end{equation}
In Appendix \ref{apx:eigensystem} we list all possible eigenvalues and eigenvectors of the covariance matrix.
The maximal QFI is found to be $F_Q= 4{\rm max}(\lambda_A,\lambda_B)$, and 
\begin{align}
\label{eq:eigenvalue1}
\lambda_A & =\Gamma_{55} \text{\ \ with\ \ } \hat{\Lambda}^{(A)}_{\mathbf{n}} = (\hat{D}_{xy} + \alpha \hat{Q}_{xy})/\sqrt{1 + \alpha^2}, \\
\lambda_B & = \left( \Gamma_{11} + \Gamma_{22} + \sqrt{4\Gamma_{12}^2 + (\Gamma_{11} - \Gamma_{22})^2} \right)/2 \text{\ \ with} \nonumber\\
\label{eq:eigenvalue2}
\hat{\Lambda}^{(B)}_{\mathbf{n}} & = \left[ (\hat{J}_x + \gamma \hat{Q}_{zx}) + \alpha (\hat{J}_y + \gamma \hat{Q}_{yz})\right]/\mathcal{N},
\end{align}
where $\gamma =(\Gamma_{22}-\Gamma_{11}+\sqrt{4\Gamma_{12}^2 + (\Gamma_{11} - \Gamma_{22})^2})/(2\Gamma_{12})$, $\mathcal{N}=\sqrt{(1 + \alpha^2)(1 + \gamma ^2)}$ and $\alpha$ is any real number. 
We found that the diagram of the QFI consists of two regions $A$ and $B$ characterized by the interferometers $\hat{\Lambda}_{\mathbf{n}}^{(A)}$ and $\hat{\Lambda}_{\mathbf{n}}^{(B)}$, respectively. The border between the regions is defined as $\lambda_A(q_t) = \lambda_{B}(q_t)$, where $q_t$ is the threshold point.

The operators $\hat{\Lambda}^{(A)}_{\mathbf{n}}$ and $\hat{\Lambda}^{(B)}_{\mathbf{n}}$ define the interferometers which would optimally  employ the thermal state in the regions $A$ and $B$, respectively.
As in the scheme discussed in Sec. \ref{sec:Three-mode interferometry}, the optimal interferometer which rotates the input state with the operator $e^{ -i\theta \hat{\Lambda}^{(A/B)}_{\mathbf{n}} }$, e.g. for $\alpha=0$, is composed of two pulses $e^{\pm i\pi \tilde{J}_y/2}$ and the phase imprinting physical term $e^{-i\theta \hat{J}_z}$. The operator $\tilde{J}_y$ used in the interferometric scheme description takes different forms in both regions as described below.

\subsection{Zero variance of magnetization, $\sigma=0$.}

When the variance of magnetization tends to zero, the state \eqref{eq:rho_state} becomes the thermal state
\begin{displaymath}
\hat\rho_M=\frac{e^{-\beta \hat{\mathcal{H}}}}{Z_M}=\sum_n p_n |n\rangle_M \, {}_M\langle n|,
\end{displaymath}
where $p_n=e^{-\beta \epsilon_{n,M} }/\mathcal{Z}_{M}$, $\epsilon_{n,M}$ is the dimensionless energy spectrum, $|n\rangle_M$ are eigenstates of the Hamiltonian in the subspace of fixed magnetization $M$.

The elements of the covariance matrix necessary to compute the QFI are given by
	\begin{eqnarray}
	\Gamma_{11}&\approx& \langle \No (\hat{N} - \No)\rangle +\frac{1}{2} \bb{\langle \hat{a}_0^2 \hat{a}_{-1}^{\dagger} \hat{a}_1^{\dagger}\rangle +c.c} , \label{eq:averages11}\\
	\Gamma_{22}&\approx&  \langle \No (\hat{N} - \No)\rangle -\frac{1}{2} \bb{\langle \hat{a}_0^2 \hat{a}_{-1}^{\dagger} \hat{a}_1
^{\dagger}\rangle + c.c. },\\
	\Gamma_{12}&\approx&  M\langle \No \rangle,\\
	\Gamma_{55}&\approx& \langle(\hat{N}-\No)^2-M^2\rangle \label{eq:averages55},
	\end{eqnarray}
where $\hat{a}_{m_F}$ is the bosonic anihilation operator of the atom in the $m_F$ component. The elements 
\eqref{eq:averages11}-\eqref{eq:averages55} are given only up to the dominant orders.

\subsubsection{The ground state: $T=0$\label{sec:T0}.}

Let us start with the simplest case, the high magnetic field limit (HMFL), when the interaction term in the Hamiltonian can be neglected.
In the HMFL one has $\hat{\mathcal{H}}/\tilde{c}\approx - q\hat{N}_{0}$ and all eigenstates of the Hamiltonian are simply given by the Fock states $\ket{N_{+1}, N_0, N_{-1}}$ with the eigenenergies $E ({N_0}) \approx - q N_0$. 
For example, when $q<0$ the ground state is $|(N+M)/2, 0, (N-M)/2\rangle$ if $M$ and $N$ have the same parity. This state lies in the region $A$, namely the QFI is given by $4\lambda^{(0)}_A=4 \Delta^2 \hat{D}_{xy} = 2 N^2(1-m^2 )$, where $m=|M|/N$ is the positive fractional magnetization. 
In the region $B$ occurring for $q>0$ in order to minimize energy one has to maximize occupation of the $m_F=0$ component, hence the ground state is $|M, N-M,0\rangle$ for positive $M$ with the QFI equal to $4\lambda^{(0)}_B=8N^2m(1-m)$. 
In the non-interacting case ($c_2=0$) the border between the regions occurs at $q_t = 0$ where abrupt change of the QFI takes place as a result of transition between $\lambda_A$ and $\lambda_B$.

In both regions $A$ and $B$ one of the $m_F$ states is not occupied, hence the ground states can be mapped onto a two-mode system and analyzed in an appropriate $SU(2)$ subalgebra spanned by $\{\tilde{J}_x,\tilde{J}_y,\tilde{J}_z\}$. 
In the region $A$ the corresponding axes are $\tilde{J}_x=\hat{D}_{xy}, \tilde{J}_y=\hat{Q}_{xy}$, $\tilde{J}_z = \hat{J}_z$, while in the region $B$ they are $\tilde{J}_x=(\hat{J}_x +\tilde{\gamma}\hat{Q}_{zx})/\sqrt{2}$, $\tilde{J}_y = (\hat{J}_y +\tilde{\gamma}\hat{Q}_{yz})/\sqrt{2}$, $\tilde{J}_z = (\hat{J}_z + \tilde{\gamma}\sqrt{3} \hat{Y})/2$ with $\tilde{\gamma}={\rm sign}(M)$. The optimal value of the QFI is determined by e.g. fluctuations of $\tilde{J}_x$.
The Reader could realize that in the region $B$, the rotation around $\tilde{J}_y$ do not map the optimal axis $\tilde{J}_x$ to $\hat{J}_z$ as was required in Sec. \ref{sec:Three-mode interferometry}. Fortunately, because $\hat{Y}$ and $\hat{J}_z$ commutes, both operators and their linear combinations should be equally good for the interferometric purposes, once it will end up with the measurements of occupations of the $m_F$ Zeeman states.

\begin{figure}[t]
\begin{picture}(0,70)
\put(-110,0){\includegraphics[width=0.19\linewidth]{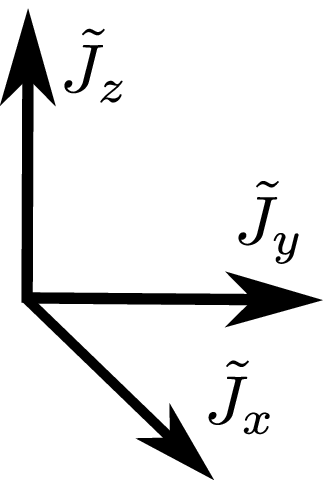}}
\put(-47,0){\includegraphics[width=0.32\linewidth,height=.32\linewidth]{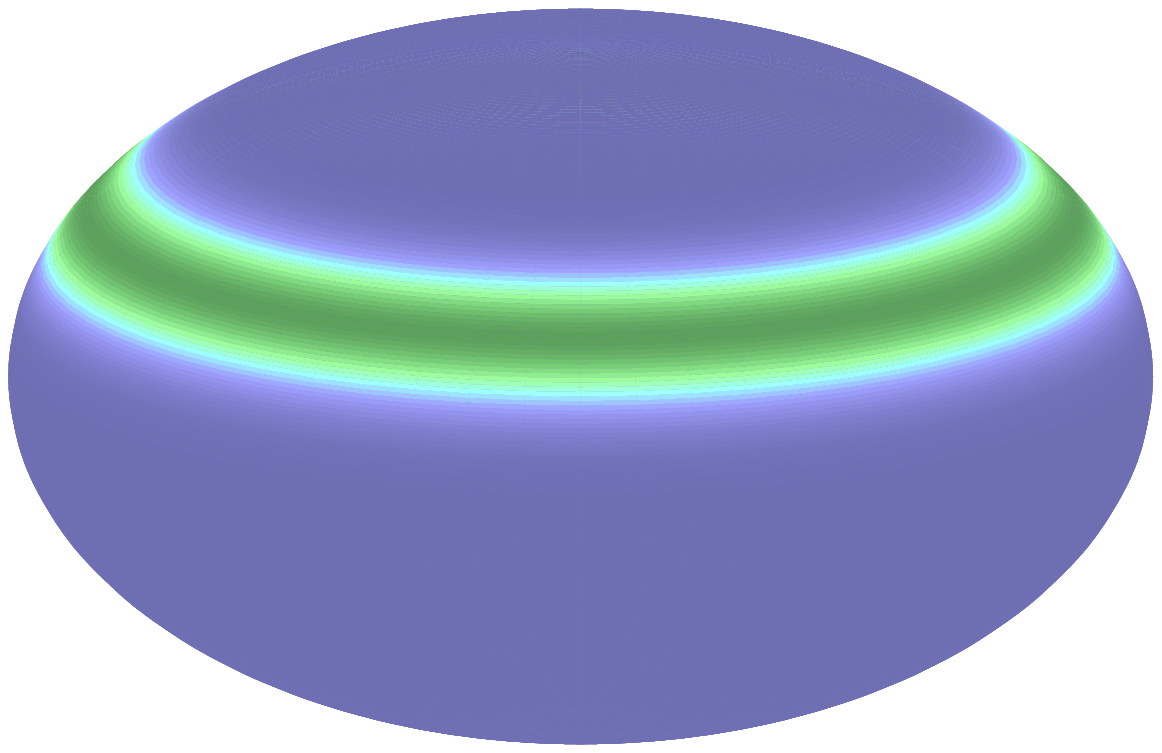}}
\put(-50,65){(a)}
\put(46,0){\includegraphics[width=0.32\linewidth,height=.32\linewidth]{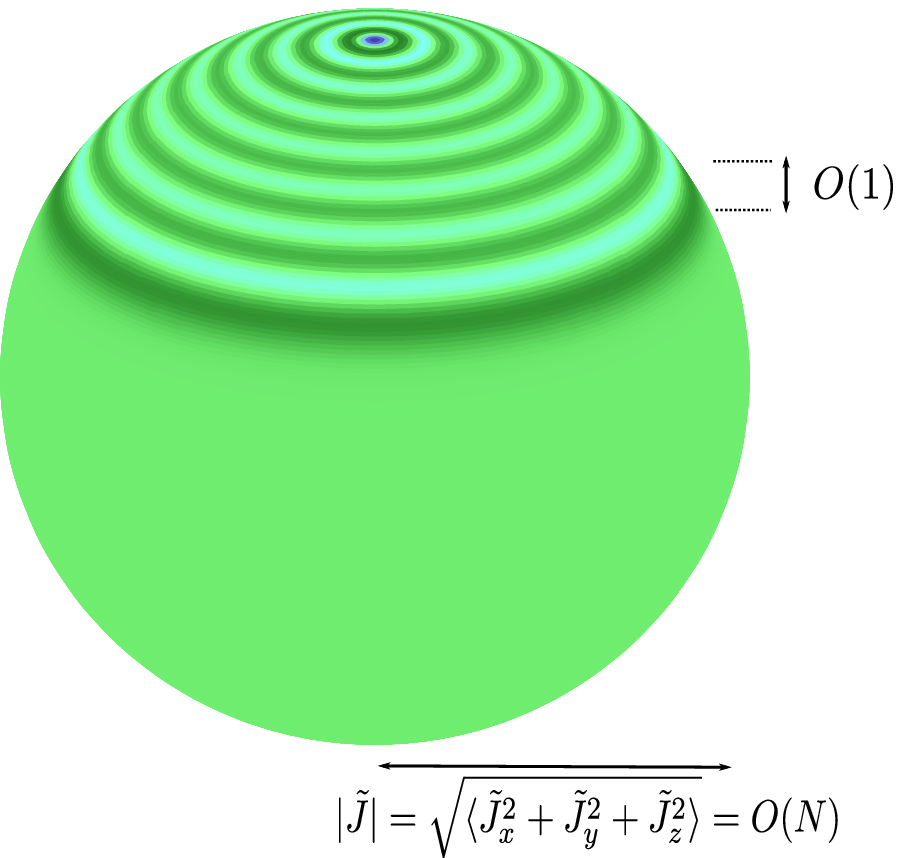}}
\put(37,65){(b)}
\end{picture}
\caption{Illustration of (a) the Husimi and (b) the Wigner quasi-distribution functions for eigenstates of the system: an example for the state $|(N+M)/2, (N-M)/2\rangle$ with $N=60, \, M=N/2$.
The maximum of both the Husimi and Wigner distributions is centered around $\langle \tilde{J}_z \rangle$, which is $M$ in the region $A$ and $2M-N$ in the region $B$.
}
\label{fig:fig0}
\end{figure}
The ground state of the system in the HMFL may be depicted with the Wigner and Husimi quasi-distribution functions in the $SU(2)$ subalgebra spanned by $\{\tilde{J}_x,\tilde{J}_y,\tilde{J}_z\}$, as illustrated in Fig.\ref{fig:fig0}.
The Wigner function of the ground state consists of latitude rings of width $1$. The width of rings compared to the Bloch sphere radius of length $N$ is a signature of the Heisenberg scaling of the QFI \cite{SmerziTreutlein}. 
Heuristically, this implies very fast changes of the state due to rotation around any axis in the $x-y$ plane of the corresponding Bloch sphere, giving rise to the Heisenberg scaling of the QFI. 

\begin{figure}[]
\includegraphics[width=\linewidth]{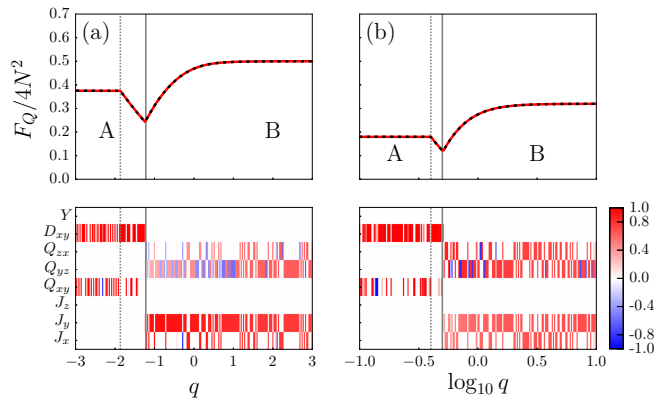}
\caption{Top row: 
QFI as a function of $q$ in the ground state of the system for $\sigma\to 0$, $N = 10^4$ and (a) $c_2<0$, $m=0.5$, (b) $c_2>0$, $m=0.8$. 
Solid red lines denote numerical results from exact diagonalization of the Hamiltonian in the Fock state basis. 
Dashed black lines denote mean-field results, as explained in the text. 
Dashed vertical gray lines mark the position of the critical points $q^{(A)}_c= (1-\sqrt{1-m^2})$ for $c_2>0$, and $q^{(F)}_c= -(1+\sqrt{1-m^2})$ for $c_2<0$ \cite{Zhang2003}.
Solid vertical gray lines mark the location of the threshold points $q_{t\pm}$ which separate the regions $A$ and $B$. 
Bottom row: eigenvectors corresponding to the largest eigenvalue of the covariance matrix as a function of $q$, demonstrating the optimal interferometer and confirming the validity of $\hat\Lambda_{\mathbf{n}}^{(A,B)}$.}
\label{fig:fig1}
\end{figure}

In the top row of Fig.~\ref{fig:fig1} we show an example of the QFI variation with respect to the parameter $q$ for both antiferromagnetic and ferromagnetic interactions obtained by exact numerical diagonalization of the Hamiltonian \eqref{eq:single_mode_ham} in the Fock state basis. 
The diagram of the QFI consists of two regions $A$ and $B$, as in the HMFL, but interactions shift the threshold point $q_t$ to positive or negative values of $q$ depending on the interaction sign. Moreover, when the interaction part dominates and $m \geqslant 0.5$, the QFI is reduced with respect to the HMFL. On the other hand, if $m < 0.5$, one can observe a local maximum in the region $B$ (not presented in Fig.~\ref{fig:fig1}).

The numerical results can be explained by the mean-field approach, which technically means the following substitutions
$\langle \No \rangle\to N \rho_0$, 
$\langle \No^2 \rangle\to N^2 \rho_0^2$ and 
$\bb{\langle \hat{a}_0^2 \hat{a}_{-1}^{\dagger} \hat{a}_1^{\dagger}\rangle +c.c} \to 2N^2\rho_0\sqrt{\rho_1\rho_{-1}}\cos(\theta)$, where $\theta=0$ for ferromagnetic and $\theta = \pi$ for antiferromagnetic interactions, in Eqs.(\ref{eq:averages11})-(\ref{eq:averages55}).
The fraction of atoms $\rho_{m_F}$ in the $m_F$th Zeeman component are then derived by minimization of the mean-field energy functional in the subspace of fixed magnetization \cite{Zhang2003}. 
The QFI can be expressed in terms of a single parameter $\rho_0$ as $\rho_{\pm 1}=(1\pm \rho_0+m)/2$.
In general, $\rho_0\in(0,1-m)$ and is a function of $q$. 
Finally, independently on interaction sign in the large atom number limit the QFI is given by
\begin{equation}\label{eq:QFIMF}
\frac{F_Q}{4 N^2}=\left\{
\begin{array}{ccc}
\frac{1}{2} \left( (1-\rho_0)^2 - m^2 \right) & \,\,\,\,& {\rm for} \, \rho_0\le \rho_t, \\
2 \rho_0 (1-\rho_0) & \,\,\,\,&  {\rm for}\, \rho_0 \ge \rho_t ,
\end{array}
\right.
\end{equation}
where $\rho_t=(3-\sqrt{4+5m^2})/5$ is the fraction of atoms in the $m_F=0$ component at the threshold point. 
From the numerical study of the relation $\rho_0(q_t) = \rho_t$ we found the approximated formulas for the threshold point $q_t$: $q_{t-}\approx-1.2$ for $c_2<0$ and $q_{t+} \approx 0.8 m^2$ for $c_2>0$.
The above formulas agree within $1 \%$ with the threshold points obtained by exact numerical diagonalization.
In the bottom row of Fig.~\ref{fig:fig1} we also show eigenvectors corresponding to the maximal eigenvalue of the covariance matrix. Numerical results confirm that for $q < q_{t\pm}$ the eigenvector and thus the optimal interferometer is $\hat{\Lambda}_{\mathbf n}^{(A)}$, while for $q > q_{t\pm}$ it is $\hat{\Lambda}_{\mathbf n}^{(B)}$.

We stress that the mean-field approach has been used only as a computational tool to estimate the covariance matrix elements given by Eqs. \eqref{eq:averages11}-\eqref{eq:averages55}.

\begin{figure}[]
\includegraphics[width=\linewidth]{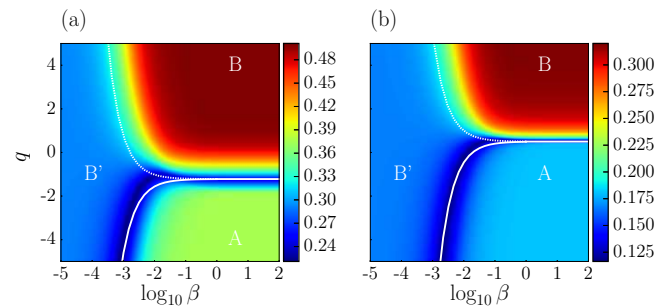}
\caption{
Exact numerical results for $F_Q/4N^2$ as a function of the temperature inverse $\beta$ and $q$ for $N = 10^3$ (a) $c_2<0$, $m=0.5$, (b) $c_2>0$, $m=0.8$.
The solid white line is the border between the $A$ and $B$, or $A$ and $B'$, regions which are approximated by $q_t(\beta)+q_{t-}$ for ferromagnetic interactions and $q_t(\beta)+q_{t+}$ for antiferromagnetic ones.
The dashed white line is the border between the $B$ and $B'$ regions approximated by $q_{BB'}(\beta)+q_{t-}$ in (a) and $q_{BB'}(\beta)+q_{t+}$ in (b).}
\label{fig:fig2}
\end{figure}

\subsubsection{The thermal state: $T\ne0$.}

In the simplest case of the HMFL the eigenstates of the Hamiltonian are Fock states and 
we used the following parametrization $|n\rangle_M\to|n,N+M-2n,n-M  \rangle_M$ which implies
that $p_n\to e^{q\beta (N+M-2n)}/Z_M$. The thermal state of the system in the sector of fixed magnetization becomes then 
$\hat{\rho}_M=\sum_{n=n_{min}}^{n_{max}} p_n |n\rangle_M \, {}_M\langle n|$ with $n_{min}={\rm Max}[0,M/2,M]$ and $n_{max}={\rm Min}[N,(N+M)/2,N+M]$.

In the simplest case, the two eigenvalues of the covariance matrix $\lambda_{A,B}$ which determine the QFI can be derived analytically, and in the large atom number limit they are
\begin{align}
\label{eq:lambdaA}
\frac{\lambda^{(\beta)}_A}{N^2} & \simeq 
\frac{1+\tilde{\beta} m} {\tilde{\beta}^2\left(1-{\rm e}^{-\tilde{\beta} (1-m)}\right)} 
- \frac{1+\tilde{\beta} +\frac{\tilde{\beta}^2}{2} -\frac{(\tilde{\beta} m)^2}{2}}{\tilde{\beta}^2 \left({\rm e}^{\tilde{\beta} (1-m)}-1\right)},\\
\label{eq:lambdaB}
\frac{\lambda^{(\beta)}_B}{N^2} & \simeq
\frac{2(1-m) (1+ \tilde{\beta} m)}{\tilde{\beta}\left( 1-{\rm e}^{-\tilde{\beta} (1-m)}\right)} - \frac{2+\tilde{\beta} + \tilde{\beta} m}{\tilde{\beta}^2},
\end{align}
where $\tilde{\beta}=q\beta N$. 
The threshold point $q_t$ depends on temperature now, and one can evaluate it by comparing the leading terms in the Taylor expansions of $\lambda^{(\beta)}_{A}$ and $\lambda^{(\beta)}_{B}$ obtaining $q_t(\beta) \approx -12 m/[(1-m)(1+5m)N\beta]$. 
In the high temperature limit, when $\beta\to 0$, the largest eigenvalue of the covariance matrix is $\lambda^{(\beta)}_B$ independently on the value of $q$.
Apart from the regions $A$ and $B$ we introduce the third one, denoted $B'$, which appears for any value of $q$ in the high temperature limit $\beta\to 0$ and in which $\hat\Lambda_{\mathbf{n}}^{(B)}$ remains still the optimal interferometer.

The border between the $B$ and $B'$ regions can be found from an inflection point of (\ref{eq:lambdaB}) which we approximated by $q_{BB'}(\beta) \simeq [\beta N(1-m)]^{-1}$.
We found that in each of the three regions, away from the borders, the QFI is practically constant and of the order of $O(N^2)$.
Consequently, the landscape of the QFI in the $\beta-q$ plane has the form of three plateaus with the universal values $\lambda^{(0)}_A$, $\lambda^{(0)}_B$, $\lambda_{B'}=N^2 (1-m)(1+5m)/6$ in the regions $A$, $B$ and $B'$, respectively.

The interacting system can only be analyzed numerically, except for the case $q=0$. 
In Fig.~\ref{fig:fig2} we show the exact numerical results for the QFI as a function of $q$ and $\beta$ where $c_2\ne 0$. 
Indeed, in addition to the $A$ and $B$ regions the third region $B'$ for $\beta\to 0$ is also preserved by interactions. 
Deeply inside the $A$, $B$ and $B'$ regions the QFI matches results derived above in the HMFL, corresponding to zero temperature results in $A$ and $B$, and to the high temperature result in $B'$
\footnote{In region $B'$, we have $p_n\to 1$ in Eq.~(\ref{eq:covariance}), for all $n$, and based on the orthonormal properties of the Hamiltonian (\ref{eq:single_mode_ham}) eigenvectors one can also show that $F_Q \simeq N^2 (1-m)(1+5m)/6$.}. 
The borders between particular regions, marked by white lines in Fig.~\ref{fig:fig2}, follow the HMFL results with vertical translation $q_{t\pm}$ correcting for the effect of interactions.  
In the all three regions the QFI has the Heisenberg scaling, as we calculated in the HMFL, irrespective of the interaction strength. 
The HMFL analysis holds, because for macroscopic magnetization all eigenstates of the Hamiltonian remain indeed close to the Fock states.

The Heisenberg scaling for $\sigma \to 0$ may be deduced directly from the formulas for the density matrix eigenvalues and the covariance matrix elements \eqref{eq:averages11}-\eqref{eq:averages55}. One can find that the sum of two eigenvalues:
\begin{eqnarray}
\lambda_B +  &\lambda_A& = \frac{1}{2}\bb{\Gamma_{11}+\Gamma_{22} + \sqrt{4\Gamma_{12}^2+\bb{\Gamma_{11}-\Gamma_{22}}^2}} + \Gamma_{55}\nonumber\\
 &\geq& \frac{1}{2}\bb{\Gamma_{11}+\Gamma_{22}} + \Gamma_{12} + \Gamma_{55} \\
 & =& \langle (\hat{N}-M) (\hat{N}- N_0+M) \rangle \geq (\langle\hat{N}\rangle -M )M \nonumber,
\end{eqnarray}
has to be large. In the case of fixed and macroscopic magnetization, i.e. $M=O(N)$ and $N-M = O(N)$,  it means that at least $\lambda_A$ or $\lambda_B$ has to scale as $\langle\hat{N}\rangle^2$.
Consequently, the QFI has to have the Heisenberg scaling for nonzero temperature and even if the total number of atoms fluctuates. 

\subsection{Non-zero variance of magnetization, $\sigma\ne 0$.}

\begin{figure}
\includegraphics[width=\linewidth]{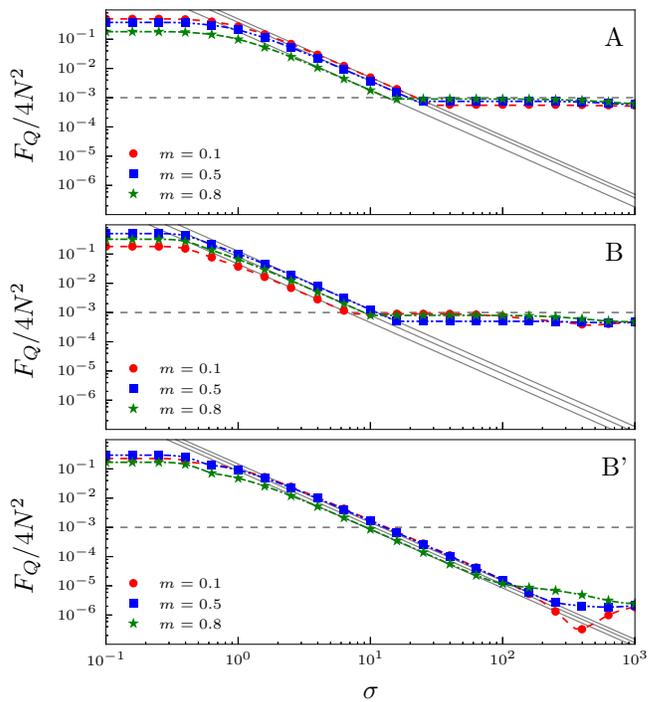}
\caption{(Color online) The QFI as a function of $\sigma$ deeply inside (a) the region $A$ for $q=-10$, $\beta=100$, (b) the region $B$ for $q=10$, $\beta=100$ and (c) the region $B'$ for $q=10$, $\beta=10^{-5}$, with $N=10^3$. 
Symbols are exact numerical results. 
Color lines denote numerical results in the HMFL. 
Gray solid lines denote decay rates, as explained in the text. 
Gray dashed horizontal lines show the SQL for $N=10^3$.}
\label{fig:fig3}
\end{figure}


As in the previous subsection, we open discussion by analysis in the HMFL for which decay rates of the QFI can be derived analytically. 
Intuitively, one may expect that the variance of magnetization larger than $1$ ($\Delta^2M=\sigma^2>1$) smears out narrow structures in the corresponding quasi-distributions illustrated in Fig.\ref{fig:fig0}, resulting in decrease of the QFI.
Lets concentrate on the region $A$ in which the QFI is determined by $\hat{\Lambda}^{(A)}_{\mathbf{n}}=\hat{D}_{xy}$ (taking $\alpha=0$ in Eq.~(\ref{eq:eigenvalue1})). The QFI is just $F_Q(\sigma)=4\lambda_A(\sigma)$ with
\begin{align}
\lambda_A(\sigma)&=\sum_{M=-N}^{N} w_M \left(N+\frac{1}{2}\left( N^2-M^2\right) \right) \nonumber\\
&- 2 \sum_{M=-N}^{N} \frac{w_M w_{M+2}}{w_M +w_{M+2}}\frac{N-M}{2}\left(\frac{N+M}{2} +1 \right)\nonumber\\
&- 2 \sum_{M=-N}^{N}\frac{w_M w_{M-2}}{w_M +w_{M-2}}\frac{N+M}{2}\left(\frac{N-M}{2} +1 \right).\label{eq:FQsigmaA}
\end{align}
The dominant term in the Taylor expansion of (\ref{eq:FQsigmaA}) around small $\sigma^{-2}$ (but with fixed N and $\bar{M}$) is 
$\sigma^{-2} N^2(1-\bar{m}^2 )/2$
which is nothing else than the QFI for $T=0$ multiplied by $\sigma^{-2}$. Hence, $\lambda_A(\sigma) \approx \sigma^{-2} N^2(1-\bar{m}^2 )/2  = \sigma^{-2} \lambda^{(0)}_A|_{m=\bar{m}}$ with $\bar{m}=\bar{M}/N$.
The same calculation can be perform in the region $B$ for which $\lambda_B(\sigma)\approx\sigma^{-2} \lambda^{(0)}_B|_{m=\bar{m}}/4$. 
Eventually for large $\sigma$ the distribution $w_M$ deviates from the Gaussian and converges to the uniform distribution $w_M\to1/(2N)$. In this limit $F_Q\to 2N$, which is two times smaller than the SQL for the QFI.
Similar analysis can be done deeply inside the region $B'$, which gives $\lambda_{B'}(\sigma)\approx \sigma^{-2} \lambda_{B'}|_{m=\bar{m}}/2$. 

In Fig. \ref{fig:fig3} we show exact numerical results including interactions for parameters deeply inside each of the regions $A$, $B$, and $B'$, and we compare them to the formulas based on the analysis in the HMFL; a good agreement is observed.
In general, based on the Taylor expansion of the factors $(v_k - v_l)^2/(v_k + v_l)$ with $v_k=p_n w_M$ in (\ref{eq:covariance}), one can expect decay rates of the QFI to be proportional to $\sigma^{-2}$ for any interaction strengths and temperatures. 
Hence, the minimal required resolution to beat the SQL is $\sigma < \sqrt{N}$, which is at the edge of current experiments with spin-1 gases \cite{PhysRevA.93.023614}.

\section{Conclusions}

We showed that spin-1 Bose condensates with macroscopic magnetization of reduced variance at thermal equilibrium (\ref{eq:rho_state}) are good candidates for atomic interferometry. Precisely, we investigated the diagram of the QFI in the $q-\beta$ plane finding three regions: (i) $A$ and $B$ which reflect the ground state structure in the low temperature limit and (ii) $B'$ in the high temperature limit.
The borders between the regions have been estimated from the results in the HMFL corrected by a shift obtained within the mean field approach.
We showed that the QFI in all three regions reaches the Heisenberg scaling for fixed and macroscopic magnetizations even at finite temperature.
The QFI decreeses for $\sigma >1$, but the SQL can be still overcome when the variance of magnetization $\sigma^2$ is smaller than $N$. 
In principle, an experimental preparation of the states we investigated does not require any additional steps. Once the system with macroscopic magnetization of reduced variance at given $q$ reaches thermal equilibrium, it would have a large QFI value.

Our results demonstrate yet another possibility to overcome the SQL by reduction of the variance of some observable, which is magnetization in our system. We expect that the SQL may be beaten, whenever the Hilbert space is restricted by a conservation law and experimental techniques to a small subspace only.

We acknowledge discussions with F. Gerbier, R. Demkowicz-Dobrza\'nski and R. Augusiak. This work was supported by the Polish National Science Center Grants DEC-2015/18/E/ST2/00760 and 2014/13/D/ST2/01883.

\appendix

\section{\label{apx:paramters} Experimental parameters}
In the following we present examples of physical parameters that are based on the standard Thomas-Fermi approximation (TFA). In the TFA one has $|\phi(\mathbf{r})|^2=\mu \omega^2(r_{TF}^2 - r^2)/(2 c_0 N)$ 
with the TF radius $r_{TF}^5=15N c_0/(4 \pi \mu \omega^2)$. The energy unit used in the Letter expressed in terms of the TF radius is 
\begin{equation}
\tilde{c}=\frac{2}{7}\frac{|c_2|}{c_0}\left(\frac{r_{TF}}{a_{ho}} \right)^2\hbar \omega,
\end{equation}
where $a_{ho}=\sqrt{\hbar/\mu\omega}$. The associated coefficient $q=Q/\tilde{c}$ in the Zeeman energy is determined by $Q=(\mu_B {\cal B})^2/(4 E_{hf})$,
that is, $Q\approx h({\cal B}/G)^2 \,277$Hz for sodium and $Q\approx h({\cal B}/{\rm G})^2 \,72$Hz for rubidium atoms.

Particular parameters calculated in SI units for $N=10^3$ atoms placed in the symmetric 3D trap with the frequency $\omega/(2\pi)=300$Hz are:\\
(i) {\bf Sodium-23:} $\tilde{c}/\hbar\approx 70$Hz; $q=1$ gives the magnetic field ${\cal B}\approx0.2$G; $\beta=1$ gives the temperature $T\approx0.53$nK with $k_BT/\hbar\omega\approx0.037$.
In addition, $q\in(0,5)$ corresponds to ${\cal B}\in(0,0.45)$G, while $\beta\in(10^{-3},10^2)$ to $T\in(533,5.3\times10^{-3})$nK. \\
(ii) {\bf Rubidium-87:}
$\tilde{c}/\hbar\approx 17$Hz; $q=1$ gives the magnetic field ${\cal B}\approx0.2$G; $\beta=1$ gives the temperature $T\approx0.13$nK with $k_BT/\hbar\omega\approx0.01$.
In addition, $q\in(0,5)$ corresponds to ${\cal B}\in(0,0.44)$G, while $\beta\in(10^{-3},10^2)$ to $T\in(133,1.3\times10^{-3})$nK. \\
Other parameters can be taken from~\cite{{PhysRevE.78.066704},{PhysRevA.75.023617}}.

\section{\label{apx:generators} $SU(3)$ Lie algebra generators}
A bosonic Lie algebra is constructed from the matrix Schwinger representation:
	\begin{equation}
	   \hat{\Lambda}_{\mu} = \sum\limits_{m,n = -1,0,+1} \left(\Lambda_{\mu} \right)^{m}_{n}\hat{a}^{\dagger}_{m}\hat{a}_{n},
	\end{equation}
where $\left(\Lambda_{\mu} \right)^{m}_{n}$ denotes the $m$th row and $n$th column of the matrix $\Lambda_{\mu}$.
The matrix representation of eight hermitian generators of the $SU(3)$ Lie algebra is given below
	\begin{align}\label{eq:generators}
	   \begin{array}{cc}
	      J_x = \frac{1}{\sqrt{2}} \left( 
			    \begin{array}{ccc}
			      0 & 1 & 0\\
			      1 & 0 & 1\\
			      0 & 1 & 0
			    \end{array}\right),
		  &
		  J_y = \frac{i}{\sqrt{2}} \left( 
		  			    \begin{array}{ccc}
		  			      0 & -1 & 0\\
		  			      1 & 0 & -1\\
		  			      0 & 1 & 0
		  			    \end{array}\right),
		  \\[6mm]
		  J_z = \left( 
			  		  	\begin{array}{ccc}
		  		  		  1 & 0 & 0\\
		  		  		  0 & 0 & 0\\
		  		  		  0 & 0 & -1
		  		  		\end{array}\right),
		  &
		  Q_{xy} = i \left( 
		  			  		 \begin{array}{ccc}
		  		  		  		  0 & 0 & -1\\
		  		  		  		  0 & 0 & 0\\
		  		  		  		  1 & 0 & 0
		  		  		  		\end{array}\right),	
		  \\[6mm]
		  Q_{yz} = \frac{i}{\sqrt{2}} \left( 
		  		  			    \begin{array}{ccc}
		  		  			      0 & -1 & 0\\
		  		  			      1 & 0 & 1\\
		  		  			      0 & -1 & 0
		  		  			    \end{array}\right),	  
		  &
		  Q_{zx} =  \frac{1}{\sqrt{2}} \left( 
		  			    \begin{array}{ccc}
		  			      0 & 1 & 0\\
		  			      1 & 0 & -1\\
		  			      0 & -1 & 0
		  			    \end{array}\right),
		  \\[6mm]
		  D_{xy} = \left( 
		  			  		  	\begin{array}{ccc}
		  		  		  		  0 & 0 & 1\\
		  		  		  		  0 & 0 & 0\\
		  		  		  		  1 & 0 & 0
		  		  		  		\end{array}\right),
		  &
		  Y =  \frac{1}{\sqrt{3}}\left( 
		  			  		  	\begin{array}{ccc}
		  		  		  		  1 & 0 & 0\\
		  		  		  		  0 & -2 & 0\\
		  		  		  		  0 & 0 & 1
		  		  		  		\end{array}\right).
\end{array}
\end{align}

\section{\label{apx:eigensystem} Eigenvalues and eigenvectors of the covariance matrix}
Below we list all possible eigenvalues and eigenvectors of the covariance matrix:
\begin{itemize}
\item[1)] $\lambda_A=\Gamma_{55}$ (double degenerate). The corresponding operator is
\[ \hat{\Lambda}^{(A)}_{\bm{n}} = \frac{(\hat{D}_{xy} + \alpha \hat{Q}_{xy})}{\sqrt{1 + \alpha^2}}.\]
\item[2)] $\lambda_B=(\Gamma_{11} + \Gamma_{22} + \sqrt{4\Gamma_{12}^2 + (\Gamma_{11} - \Gamma_{22})^2})/2$ (double degenerate).
The corresponding operator is
\[ \hat{\Lambda}^{(B)}_{\bm{n}} = \frac{\left[ (\hat{J}_x + \gamma\hat{Q}_{zx}) + \alpha (\hat{J}_y + \gamma\hat{Q}_{yz})\right]}{\sqrt{(1 + \alpha^2)(1 + \gamma^2)}}.\]
\item[3)] $\lambda_C = \Gamma_{77}$.
The corresponding operator is $\hat{\Lambda}^{(C)}_{\bm{n}} = \hat{Y}$.
\item[4)] $\lambda_D=(\Gamma_{11}+\Gamma_{22} - \sqrt{4\Gamma_{12}^2 + (\Gamma_{11} - \Gamma_{22})^2})/2$ (double degenerate).
The corresponding operator is
\[ \hat{\Lambda}^{(D)}_{\bm{n}} = \frac{\left[ (\hat{J}_x + \gamma_D\hat{Q}_{zx}) + \alpha (\hat{J}_y + \gamma_D\hat{Q}_{yz})\right]}{\sqrt{(1 + \alpha^2)(1 + \gamma_D^2)}}\]
\end{itemize}
where $\gamma=(\Gamma_{22}-\Gamma_{11}+\sqrt{4\Gamma_{12}^2 + (\Gamma_{11} - \Gamma_{22})^2})/(2\Gamma_{12})$, $\gamma_D=(\Gamma_{22}-\Gamma_{11}-\sqrt{4\Gamma_{12}^2 + (\Gamma_{11} - \Gamma_{22})^2})/(2\Gamma_{12})$ and $\alpha$ is real.
Notice that $\lambda_B \geqslant \lambda_D$. We have checked that $\lambda_C=\lambda_A$ for $q\to0$ and $M\to0$, and $\lambda_C<\lambda_A$ for other cases.

\bibliography{bibliography}

\end{document}